# Hybrid Photoelectron Momentum Microscope at the Soft X-ray Beamline I09 of the Diamond Light Source

Matthias Schmitt[1,2], Deepnarayan Biswas[1], Olena Tkach[3], Olena Fedchenko[3], Jieyi Liu[1], Hans-Joachim Elmers[3], Michael Sing[2], Ralph Claessen[2], Tien-Lin Lee[1] and Gerd Schönhense[3]

[1] *Diamond Light Source Ltd., Didcot, United Kingdom*
[2] *Physikalisches Institut and Würzburg-Dresden Cluster of Excellence ct.qmat, Julius-Maximilians-Universität, D-97074 Würzburg, Germany*
[3] *Institut für Physik, Johannes Gutenberg-Universität, Mainz, Germany*

**Abstract**

Soft X-ray momentum microscopy of crystalline solids is a highly efficient approach to map the photoelectron distribution in four-dimensional (*E*, ***k***) parameter space over the entire Brillouin zone. The fixed sample geometry eliminates any modulation of the matrix element otherwise caused by changing the angle of incidence. We present a new endstation at the soft X-ray branch of beamline I09 at the Diamond Light Source, UK. The key component is a large single hemispherical spectrometer combined with a time-of-flight analyzer behind the exit slit. The photon energy ranges from hν = 105 eV to 2 keV, with circular polarization available for hν > 150 eV, allowing for circular dichroism measurements in angle-resolved photoemission (CD-ARPES). A focused and monochromatized He lamp is used for offline measurements. Under *k*-imaging conditions, energy and momentum resolution are 10.2 meV (FWHM) and 0.010 Å$^{-1}$ (base resolution 4.2 meV with smallest slits and a pass energy of 8 eV). The large angular filling of the entrance lens and hemisphere (225 mm path radius) allows *k*-field-of-view diameters > 6 Å$^{-1}$. Energy filtered X-PEEM mode using synchrotron radiation revealed a resolution of 300 nm. As examples we show 2D band mapping of bilayer graphene, 3D mapping of the Fermi surface of Cu, CD-ARPES for intercalated indenene layers and the *sp* valence bands of Cu and Au, and full-field photoelectron diffraction patterns of Ge.

# 1. Introduction

Photoelectron momentum microscopy, also termed *k*-PEEM, has proven to be a powerful method for angle-resolved photoelectron spectroscopy (ARPES) [1,2 and refs. therein]. Such instruments record $k_x$–$k_y$ images with diameters typically exceeding a full Brillouin zone. As energy filters, double- and single-hemispherical or time-of-flight (ToF) analyzers are in use. The high collection efficiency of the objective lens of a momentum microscope (MM), typically a cathode lens with extractor electrode at high voltage, allows one to observe large angular ranges up to the full 2π solid angle. MMs can cover the energy range from the near threshold region (laser-ARPES) up to the soft and even hard X-ray range.

After pioneering work with a double-hemisphere laboratory instrument by Kirschner and Tusche [3,4], several hemisphere-based MMs have been installed at synchrotron radiation sources [1], [5-7]. Experiments with ToF-MMs have been performed in 40-bunch filling mode of PETRA III (DESY, Hamburg) using soft X-rays [8] and hard X-rays, even with photoelectron spin analysis [9]. Soft X-ray momentum microscopy, in particular, is well-suited for mapping the photoelectron distribution in four-dimensional (*E*,***k***) parameter space. The fixed sample geometry is advantageous for circular dichroism (CD-ARPES) or photoelectron diffraction (PED) experiments, because it eliminates modulation of the matrix element due to a change in angle of incidence. Like the majority of synchrotron radiation sources worldwide, the Diamond Light Source (DLS) in Didcot, UK, is operated nearly exclusively at a pulse rate of approximately 500 MHz, which is not suitable for ToF-MM due to the short time gap of only 2 ns between the adjacent pulses. For such sources, the ToF technique can be accomplished in two different ways: either the pulse rate is reduced via electron-optical pulse picking using a fast deflector with GHz bandwidth in the electron path, or the energy band width is reduced by means of a dispersive bandpass filter, before the electrons enter the ToF analyzer [10].

Here we describe a novel setup of the latter type, recently installed at the soft X-ray branch of beamline I09 of the DLS. This hybrid-type ('hemisphere & ToF') microscope is based on a large single hemispherical analyzer (path radius of 225 mm) with an additional 'ToF-booster' behind the exit slit of the hemisphere. The photon energy range of this beamline covers hν = 105 eV to 2 keV, with circular polarization available for hν > 150 eV, enabling CD-ARPES measurements. For offline measurements, a focused and monochromatized He lamp is used.

The dispersive-plus-ToF hybrid mode of operation is novel for synchrotron-based experiments. Momentum-resolved recording circumvents the resolution limitation due to the $α^2$-aberration term, where α is the entrance angle into the analyzer. A base resolution of 4.2 meV (FWHM) has been obtained with small analyzer slits (200 μm) and a pass energy of 8 eV [11]. At conditions usable for synchrotron experiments ($E_{pass}$= 10 eV and 400 μm slits), we found 10 meV resolution. The transit-time spread due to different path lengths of the electrons in the analyzer is commonly considered detrimental for time-resolved experiments with hemispherical analyzers [12,13]. However, owing to the *k*-resolved detection, the spread can be easily corrected numerically.

We show examples for several application cases: X-PEEM measurements for several test samples, band mapping of double-layer graphene, 3D energy surface mapping of bulk copper [Cu(111)], CD-ARPES for intercalated indenene layers and for the *sp* valence bands of Cu and Au, and a series of full-field photoelectron diffraction patterns of Ge.

## 2. Layout of the instrument and characterization

### *2.1    The setup at the I09 beamline*

The momentum microscope is set up at on of the soft X-ray branches of beamline I09 [14] at the Diamond Light Source, providing a photon energy range of 105 – 2000 eV. The instrument is embedded in a UHV endstation consisting of a loadlock, distribution and storage chamber as well as a preparation chamber equipped with a water-cooled 4-axis manipulator with both a resistive and an electron-beam heater, an argon sputter gun, a LEED and further ports for user equipment like, e.g., evaporators. In addition, there is dedicated port for mounting a UHV suitcase to the endstation enabling *in situ* sample transfer from other preparation facilities.

Figure 1 shows the momentum microscope featuring a hexapod manipulator allowing a 5-axis sample alignment and liquid helium cooling for studies at cryogenic temperatures down to around 28 K. The main lens column features an extractor lens to which voltages of up to 20 kV can be applied, two stigmators in a Gaussian and reciprocal plane as well as different sets of field and contrast apertures. The 500 MHz filling pattern of the Diamond storage ring, resulting in a short bunch separation of only 2 ns, necessitates for time-of-flight recording a dispersive bandpass filter, here a large hemispherical analyzer (SPECS Phoibos 225). This analyzer transmits a specific energy range – depending on the pass energy and slit widths – of photoelectrons, while preserving their two-dimensional momentum distribution. The electron lenses are adjusted such that the entrance and exit slits host Gaussian images. A second lens system behind the analyzer converts the Gaussian image from the exit slit back into a momentum image in the detector plane. The base pressure of the microscope chamber is $1\times10^{-10}$ mbar and reaches $6\times10^{-11}$ mbar with liquid helium cooling.

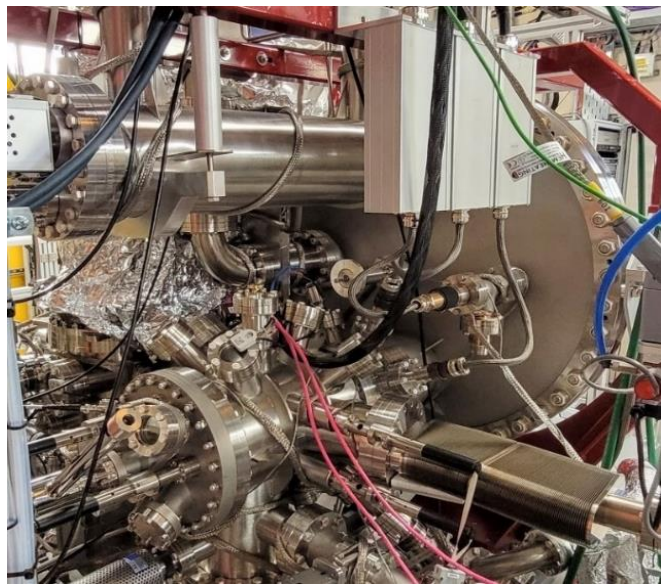

**Fig. 1.** Photo of the momentum microscope at the soft X-ray branch of I09 at Diamond Light Source: Bottom: Microscope chamber with connections for the hexapod sample manipulator on the left and the main lens column including the extractor lens built in. Right: Hemispherical analyzer (SPECS Phoibos 225) for energy bandpass pre-filtering. Top: Time-of-flight tube with built-in lens system and delayline detector on the left. The synchrotron radiation enters the microscope chamber from the right through the flat bellow, allowing for a rotation of the whole set up.

Finally, the photoelectrons travel through a low-energy drift section before being detected by a delayline detector (DLD). In the hybrid operation, this high-speed detector slices the incoming electron bunches, energy-filtered by the hemispherical analyzer, into thin time sections, determining the instrumental energy resolution. Assuming a time resolution of the DLD of about 200 ps, the gain factor can ideally be 10 for a 500 MHz pulsed source. As the rims of the energy interval cannot be used, in practice the gain factor is about 5. A key advantage of this hybrid-type MM design is its compatibility with continuous sources, in the present instrument a He lamp. At low pass energies the instrument works without ToF recording. Then only a single energy slice is measured at a time and energy series are measured sequentially. In this case (the *analyzer-only mode*) the energy resolution is determined by the pass energy and analyzer slit widths only.

The synchrotron radiation enters the microscope chamber through a flat bellow, which can be seen in Fig. 1 on the right side. This in combination with the heavy duty rotary stage carrying the multi-chamber system allows for rotation of the endstation and thus changing the angle of incidence from grazing incidence by up to 20° without the need to move the sample relatively to the extractor lens.

## 2.2 Characterization of the instrument

The entrance lens of the hemispherical analyzer contains two piezo-motor-driven arrays of field apertures and contrast apertures that allow us to limit the accepted area on the sample and the acceptance angle, respectively. For the following measurement determining the instrumental energy resolution, focused He I radiation (21.2 eV) from a monochromatized UV source, analyzer entrance and exit apertures ($d_1$, $d_2$) with a diameter of 400 µm, and a pass energy of $E_{pass}$ = 10 eV were used. Given the mean path radius of the hemisphere of $R_0$ = 225 mm, the theoretical instrumental energy resolution can be calculated as:

$$\Delta E_{theo} = E_{pass} \cdot \frac{d_1+d_2}{4 \cdot R_0} = 8.9 \text{ meV} \qquad (1)$$

As the hemisphere is used in an imaging mode, the $\alpha^2$-term does not affect the energy resolution, because in a 3D data stack it can easily be corrected numerically. For the experimental determination, a Au(111) single crystal has been freshly prepared by argon ion sputtering and subsequent thermal annealing to about 500 °C within the preparation chamber. The quality of the crystal surface was verified by observing the herring-bone reconstruction in the LEED pattern. The sample was then inserted into the LHe cooled hexapod of the MM and the Fermi cutoff was measured. This measurement was performed in the *analyzer-only mode.* By stepwise sweeping the voltage applied to the sample, the small energy window dictated by the pass energy and transmitting the electrons through the analyzer slits is shifted across the spectrum. Unlike in conventional ARPES experiments, here the sample bias is set to a positive value ($U_{sample} = E_{kin}/e$), which is advantageous because the analyzer and the ToF section are kept at low voltages.

By recording the images at a 2 mV step of the sample bias and then integrating the intensity across the $k_x$ and $k_y$ directions, we obtained the angle-integrated spectrum shown in Fig. 2a. After deconvolution with the thermal broadening at 28 K, the instrumental energy resolution was determined to be 10.2 meV and is thus in good agreement with the theoretical value.

To assess the instrumental momentum resolution, high-statistics Fermi surface mapping of the gold crystal was performed (Fig. 2b), exhibiting the well-known Shockley surface state around the Γ-point. By changing the voltages of the exit lens column, it was possible to further zoom into the region close to the Γ-point. Figure 2c shows the Fermi-energy map with an estimated lateral magnification factor of about 6. The Rashba splitting of about 0.025 Å$^{-1}$ is clearly resolved. A line profile extracted along the red dashed line is depicted in (d), revealing four distinct peaks. To quantify the instrumental resolution, the line profile was fitted with a Voigt function (shown in blue). This function incorporates both the intrinsic lifetime broadening (approximated by a Lorentzian component) and the instrumental broadening (represented by a Gaussian component). Literature values [4] for the Lorentzian widths of the inner (0.0084 Å$^{-1}$) and outer ring (0.0065 Å$^{-1}$) were employed. This way, the momentum resolution was determined to be 0.010 Å$^{-1}$, which reflects a typical value for MMs. From the measured $k_x$-$k_y$-$E_{kin}$-stack, bandmaps along different directions in k space can be extracted. In Fig. 2e the dispersion of the surface state cut along the red dashed line in (c) is shown. The Rashba splitting and the band crossing at the Γ-point around $E_B$ = 500 meV is well resolved.

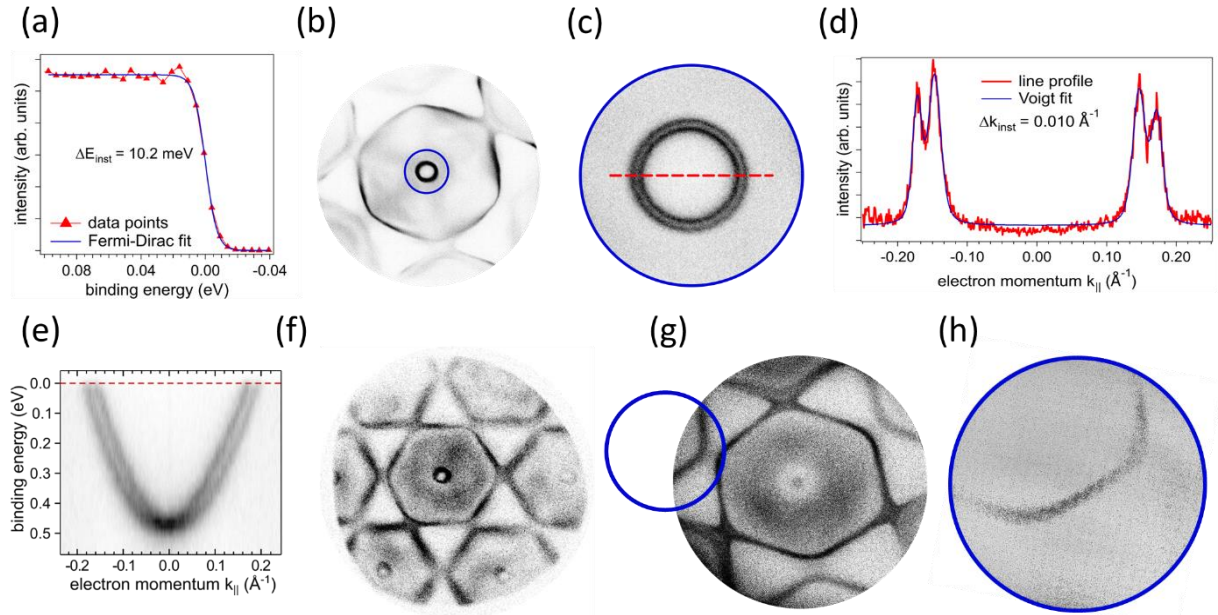

**Fig. 2** (a) Fermi edge of a Au(111) single crystal, showing an energy resolution of 10.2 meV. (b) Fermi energy map of a Au(111) crystal recorded with He I radiation (21.2 eV). (c) Center part of (b) zoomed in using electron optics, resolving the Rashba splitting of the Shockley surface state. (d) The intensity line scan extracted along the red dashed line in (c) together with a Voigt fit reveals an instrumental momentum resolution of 0.010 Å$^{-1}$. (e) Dispersion of the surface state along the red dashed line in (c) shows band splitting and the crossing at the Γ-point around $E_B$ = 500 meV. (f) Fermi energy map measured with He II radiation (40.8 eV). The photoemission horizon (emission angle range ±90°) reflects a momentum field-of-view of about 6 Å$^{-1}$. (g) Gold Fermi map measured with synchrotron radiation (120 eV). Due to the enhanced probing depth, the surface state is no longer visible. (h) Zoom-in of the off-center Fermi surface cut (blue circle in (g)), using the exit lens system and the *k*-stigmator.

Another important parameter of a momentum microscope is the zoom range of the *k*-field-of-view, i.e., the possibility between accessing momentum images in maximum detail (as in Fig. 2c) and over as many Brillouin zones (BZs) as possible. The upper limit of the latter extreme corresponds to an acceptance angle of up to ± 90° at low energies. We checked whether the full photoemission horizon can be observed using He II excitation (hν = 40.8 eV). Theoretically, the entire half-space should now result in a total field of view of 6 Å$^{-1}$, covering the first and 2/3 of every second adjacent BZ. The recorded Fermi energy pattern of Au(111) at He II is shown in Fig. 2f. Clearly, the field-of-view is larger than for He I radiation (b) and indeed shows about 2/3 of the second BZs. The intensity rim reflects the photoemission horizon, i.e., off-normal emission up to 90°.

More complicated is zooming into features which are off-center like K or M points as well as adjacent BZs. Figure 2g shows the gold Fermi map measured with synchrotron radiation (120 eV). Due to the enhanced probing depth at higher photon energies, the surface state at the Γ-point is no longer visible. The *k*-stigmator enables not only precise alignment of the momentum image to the detector, but also controlled displacement. This, in combination with the exit lens system, allows for off-center zooming with the MM. Figure 2h shows the zoomed-in adjacent BZ marked in (g).

### 2.3 Energy-filtered PEEM using synchrotron radiation

The voltages applied to the microscope lens column can be adjusted for the entrance and exit slits to host either real-space or momentum images. The latter case establishes the X-PEEM imaging mode of the hemispherical analyzer, first demonstrated in the pioneering work by Tonner [15]. At the detector, a real space image (Gaussian image) of the sample surface is formed. Figure 3a shows an X-PEEM image of a calibration sample consisting of horizontal and vertical stripe patterns of Au. The image was recorded at the Au 4*f*$_{7/2}$ core-level peak at a photon energy of 120 eV. The well resolved line patterns at the bottom have a periodicity of 2 μm. Line scans along the horizontal and vertical directions, marked by red and blue dashed lines, depicted in (b), reveal a spatial resolution in X-PEEM mode of 300 nm or better.

Figures 3c,d show X-PEEM images of a second test sample with a hierarchic checkerboard pattern of gold on silicon. In (c) the energy is set for the Au 4*f*$_{7/2}$ core level signal. Within the bright area small squares with a side length of 1 μm each are visible. In (d) the same sample region is shown with the energy set for the Si 2*p*$_{3/2}$ core level. Between (c) and (d), the contrast between the bright and dark areas is swapped. This mode facilitates the identification of distinct surface domains and structures through chemical mapping.

In threshold PEEM mode (excitation with a Hg lamp at 4.9 eV photon energy) an identical instrument showed a spatial resolution of 40 nm [11]. The contrast in those PEEM images is composed of work function and topographic contrast. Images without energy filtering can be recorded in 'straight-through mode' using an MCP-CCD camera arrangement attached to a view port behind the entrance aperture of the analyzer (Fig. 3e). In this mode, all hemisphere voltages are set to the same value, so that the beam is not deflected. This mode is suited for a rapid inspection of the sample, for searching small sample flakes and for selecting a homogeneous area on the surface.

By inserting a small field aperture it is possible to select a small region of the photon footprint, enabling μARPES. The smallest available field aperture yields an accepted area on the sample of about 5 μm.

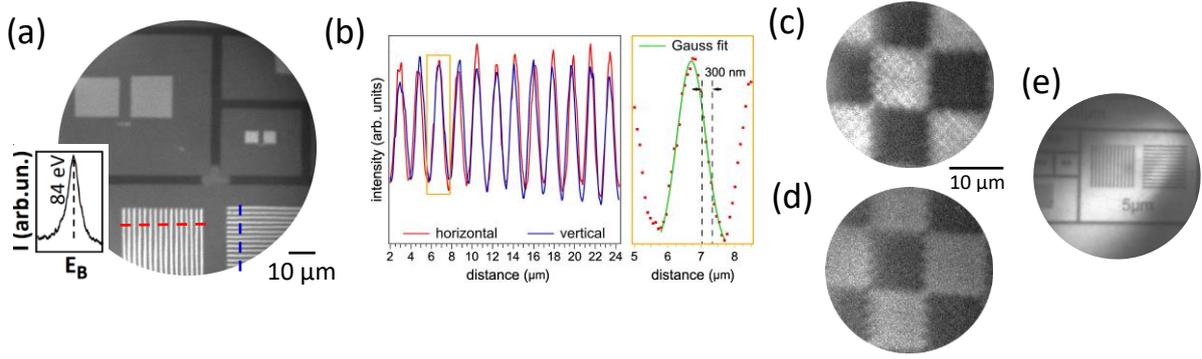

**Fig. 3.** Element specific imaging using the X-PEEM mode at hν = 120 eV. (a) X-PEEM image recorded on the Au $4f_{7/2}$ core level signal. The calibration sample consists of two fields with stripes of 2 μm spacing made from gold on silicon. (b) Line profiles extracted along the dashed lines in (a) reveal a spatial resolution of about 300 nm. (c,d) X-PEEM of a checkerboard sample with Au squares on Si; (c) and (d) recorded on the Au $4f_{7/2}$ core level and the Si $2p_{3/2}$ core level, respectively. (e) PEEM image of the same sample as in (a), but measured in the straight, non-energy-filtered branch with the He lamp. The photon footprint is clearly visible.

## 2.4 Hemisphere & ToF hybrid mode of operation

The small time gap between adjacent synchrotron radiation pulses for the 500 MHz filling pattern does not allow pure time-of-flight (ToF) based photoelectron spectroscopy. However, the hemispherical analyzer is capable of reducing the transmitted energy band to any desired value. In the hemisphere & ToF *hybrid mode* of operation the hemispherical analyzer acts as a bandpass pre-filter for the ToF analyzer. Bandpass pre-selection can enable efficient ToF momentum microscopy at a 500-MHz synchrotron.

The efficiency gain when switching from the standard 2D ($k_x,k_y$) to the 3D ($k_x,k_y,E_{kin}$) hybrid recording mode depends on several factors. This comparison must be based on identical energy resolution, which in the present setup is limited by the photon bandwidth of 50 meV. To estimate the gain factor, we start with the 10 meV resolution measured with the He lamp at $E_{pass}$= 10 eV and 400 μm entrance and exit apertures (Fig. 2a). Staying with the 400 μm apertures, $E_{pass}$= 50 eV gives a resolution of 50 meV in the 2D recording mode. Upon activating the 'ToF booster' the pass energy can be increased to $E_{pass}$= 500 eV, resulting in a bandpass of 500 meV. In the 3D recording mode the ToF analyzer resolves this band into 10 energy slices of 50 meV width. Since the edges of the time window cannot be used, we assume a realistic value of 5 usable time slices in this interval.

A second contribution to the total gain occurs at the entrance aperture. At a pass energy of 50 eV the entrance lens operates with a magnification factor of 20 between the sample surface and the entrance plane of the analyzer. This means that the 400 μm entrance aperture corresponds to a 20 μm diameter disk on the sample surface. The footprint of the photon beam has a size of 20 x 50 μm, so the image of the footprint is partially cut off by the entrance aperture. According to Liouville's theorem, the magnification varies with $E_{pass}^{-1/2}$. Thus, changing $E_{pass}$ from 50 to 500 eV, increases the accepted area on the sample corresponding to the entrance aperture of 400 μm by a factor of 10. For a homogeneously illuminated area of 60 µm diameter, the intensity gain would be a factor of 10. For the given footprint we estimate a gain factor of 3. Overall and within these conservative estimates, we can expect an efficiency gain of 5 from the number of resolved time slices and another 3 from the larger accepted area on the sample, making a total factor of 15.

The gain from the clipping effect is maximum for large photon footprints. Because it contains a Gaussian image, the analyzer entrance aperture acts as a field aperture. In a momentum microscope the field aperture plays the role of a contrast aperture for the *k*-images. For 'acceptance discs' on the sample surface with diameters between 20 and 60 µm there is no significant difference in the *k*-resolution.

For technical reasons, there was so far no synchrotron beam time allocated to using the timing mode. Therefore, the 'hemisphere & ToF' hybrid mode was tested using pulsed laser excitation. Figure 4 illustrates the recording scheme and shows two examples. The time spread of the electrons in the analyzer due to different path lengths depends on the pass energy and the transversal momentum in the dispersive direction. Here we have chosen $k_y$ and $k_x$ for the dispersive and non-dispersive directions, respectively (for details on the geometry, see Fig. 1 in [11]). Figure 4a shows a scheme of the electron distribution with the time spread along the $k_y$ direction; τ is the time of flight of the electrons after passing the hemisphere and the subsequent lens system. Energy isosurfaces such as the Fermi energy plane appear tilted in the ($k_x$,$k_y$,τ) coordinate frame. This tilt is significant and is numerically corrected by a function linear in $k_y$, as shown in (b).

This time spread is most evident in threshold photoemission, which results in a narrow energy distribution. The distribution looks horizontal in the non-dispersive direction (c) and tilted in the dispersive direction (d). The same occurs in a real spectrum, as seen in (e,f) for a thin film of Au on Re(0001), showing quantum well states. Synchrotron radiation measurements are planned.

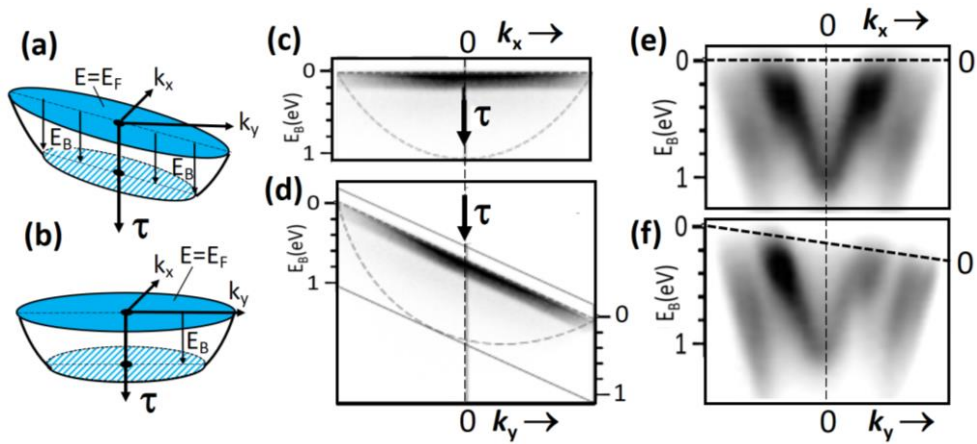

**Fig. 4.** Schematics (a,b) and example measurements of threshold photoemission (c,d) and of the quantum well states in a thin film of Au on Re(0001) (e,f) for the 'hemisphere & ToF' hybrid mode at pass energies of 200 eV (c,d) and 500 eV (e,f) using a pulsed laser. Due to the time spread of the electrons in the analyzer, the distributions appear tilted in the ($k_x$,$k_y$,τ) coordinate frame; τ time-of-flight (a). The tilt is weaker for higher pass energy, compare (d,f).

# 3. Momentum-space imaging

## 3.1 Band mapping of a 2D system

One of the key features of this instrument is the enhanced detection efficiency due to the fact that both directions, $k_x$ and $k_y$, are recorded simultaneously at a given kinetic energy. This allows for much faster investigation of, e.g., two-dimensional electron systems which specifically suffers from low count rates and thus long exposure times. Here we report on the measurements of bilayer graphene on SiC.

The sample was heated to 200 °C after inserting it into the vacuum system and then cooled down to 28 K. The beamline was set to 150 eV photon energy, resulting in high surface sensitivity and an overall energy resolution of about 50 meV. With the mirrors optimized for a best focused beam, the spot size on the sample was about 20 x 50 µm. This measurement was carried out without the ToF analyzer, and the hemisphere pass energy set to 30 eV. The 3D pattern ($k_x$, $k_y$, $E_B$) was acquired by scanning the voltage applied to the sample in steps of 20 mV and recording a ($k_x$, $k_y$)-pattern at each energy. Then all patterns were concatenated to form the final 3D ($k_x$, $k_y$, $E_B$) data array. The electron lenses were set for the *k*-field-of-view to cover a full BZ.

Figures 5a and b show two constant energy maps measured at different binding energies. Image (a) shows the hexagonal arrangement of the Dirac points at a binding energy of 360 meV. The additional faint structures (and also the weak satellite bands in Fig. 5c) result from multiple scattering of the photoelectron wave off the non-commensurate SiC and graphene lattices and the associated umklapp processes. The k-pattern at the higher binding energy of 1.5 eV in Fig. 5b reveals the expected outward dispersion of the Dirac bands [16,17]. Close inspection reveals a recurring two-band structure, consistent with the graphene bilayer. This two-band structure is most clearly visible in the $E_B$-vs-$k_\parallel$ section in Fig. 5c, showing the band dispersion along the Γ-K-M-Γ direction.

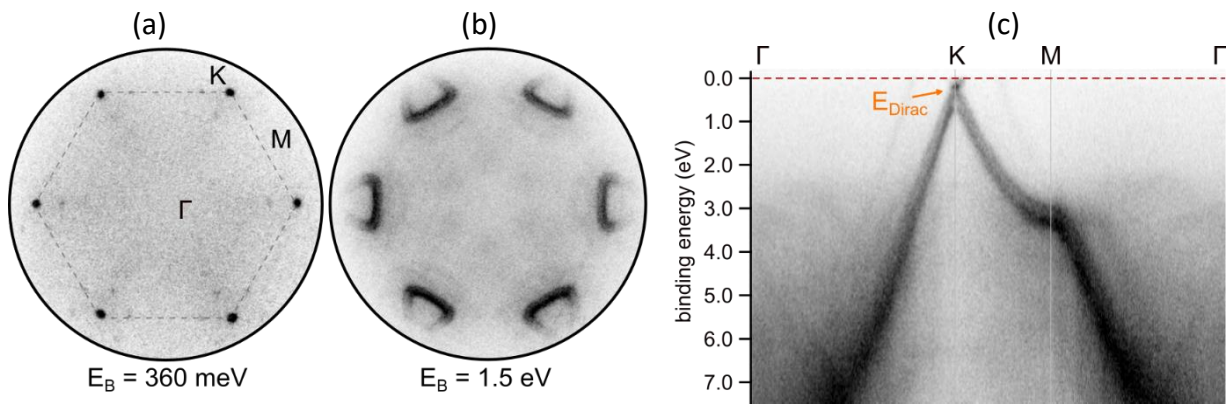

**Fig. 5.** Band mapping of a graphene bilayer on SiC at a photon energy of 150 eV (bandwidth 50 meV). (a,b) ($k_x$,$k_y$) momentum maps recorded at binding energies of 360 meV and 1.5 eV, respectively. (c) Band dispersion plot along the Γ-K-M-Γ direction, revealing the band splitting characteristic for the bilayer.

## 3.2 Energy-surface mapping in 3D k-space for a bulk system

Providing photon energies from hv = 105 eV to 2 keV, the soft X-ray branch of beamline I09 is ideally suited for probing not only surface states and adsorbates, but also the bulk electronic structure. The inelastic mean free path (IMFP) of photoelectrons in the bulk materials studied below is as low as 0.4 nm, 0.5 nm and 0.5 nm at 110 eV and rises to 3.8 nm, 2.9 nm and 2.1 nm at 2 keV for Ge, Cu and Au, respectively [18]. Hence, the lower end of the energy range is close to the minimum of the universal curve providing high surface sensitivity. Soft X-ray MM, exploiting photon energies towards the upper end, gives access to the bulk electronic structures of particularly 3D systems. Such data are of interest, e.g., in the context of transport properties. Moreover, the large IMFP allows studying buried interfaces, samples with protective coatings and deeper-lying layers.

The central application of soft X-ray MM is mapping of an electronic band structure in 3D momentum space, an efficient way of performing ARPES. Here we show as an example how full-field imaging MM is used for recording isoenergetic surfaces in $k$-space, including their circular dichroism texture (CD-ARPES). Looking directly into momentum space, we obtain $k$-patterns without the necessity to transform the data from real-space polar coordinates to $k$-space coordinates.

As an example, Fig. 6 presents mapping of the Fermi surface of Cu. The transversal coordinates $k_x$ and $k_y$ are directly accessible in the images, whereas the $k_z$ coordinate is varied by changing the photon energy. The 3D Fermi surface in Fig. 9d was concatenated from 20 $k_x$-$k_y$ patterns (dwell time 20 minutes each) like those shown in Figs. 6a,b. A $k_x$-$k_z$ section through the Fermi surface is displayed in (c), showing the region between the 'belly' (section a) and 'neck' (section b), whose cross sections in $k$-space have been measured in de Haas – van Alphen (dHvA) experiments [19,20]. In Fig. 6d the experimental Fermi surface of Cu has been truncated at the boundaries of the BZ. As expected, the MM-derived surface is qualitatively similar to the Fermi surface, constructed on the basis of the dHvA measurements (Fig. 6e).

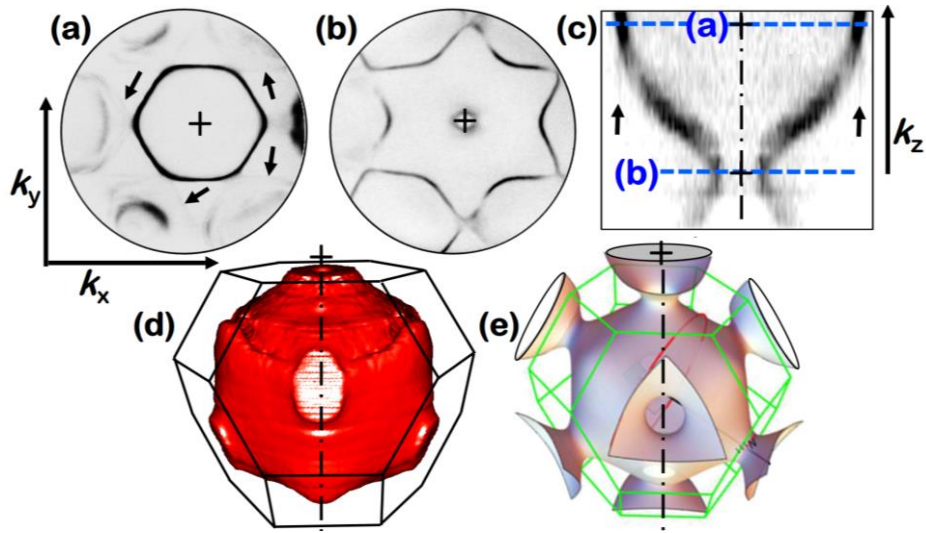

**Fig. 6.** (a,b) $k_x$-$k_y$ momentum images for Cu(111) at the Fermi energy for two different photon energies, corresponding to different $k_z$. (c) $k_x$-$k_z$ section through the 3D data array, concatenated by 20 images like (a,b). This plot shows only the innermost band, centered around the + signs in (a,b). (d) View of the experimental Fermi surface of Cu, derived from the concatenated data stack. (e) Theoretical Fermi surface of Cu, after Ref. [22].

Close inspection of the band patterns reveals that the 'belly' contour is visible as faint lines [marked by arrows in (a,c)] in all $k_x$-$k_y$ images, independent of $k_z$. This is an interesting result, which resembles the multiple-final-states signature described in [21]. We tentatively attribute this to a kind of 'density-of-states effect' in $k$-space mapping. The high density of states corresponding to the extremal orbits observed in dHvA measurements leads to faint photoemission signals even if $k_z$ 'formally' does not cut through the correct plane of the Fermi surface. The upper part of the Fermi surface, Fig. 6d, shows some distortions, since there are several interfering final states (cf. Fig. 8). These phenomena will be studied in detail alongside with photoemission calculations. The CDAD patterns corresponding to Fig. 6 are shown and discussed in Fig. 8. Four-dimensional ($E_B$,$\mathbf{k}$) data arrays (comprising many constant binding energy surfaces) would require performing the same procedure at different binding energies.

### 3.3 *Mapping of the circular dichroism texture in 2D and 3D k-space*

Circular dichroism measures the different response of a system to photon beams of the opposite helicities. In valence-band photoemission it originates from a matrix element effect in the excitation step. CD-ARPES provides a sensitive probe of wavefunctions and yields specific information on topological systems [23,24]. This effect has been termed circular dichroism in the angular distribution (CDAD) and it is described by two quantities: The *CDAD signal* is the difference of the intensities recorded with the two photon helicities $I_{RCP}$ - $I_{LCP}$; the *CDAD asymmetry* is defined as the relative quantity ($I_{RCP}$ - $I_{LCP}$)/( $I_{RCP}$ + $I_{LCP}$). In order to demonstrate the information content of CD-ARPES, we show two examples: The atomic monolayer system indenene and gold and copper as prototypical bulk metals with simple *sp* valence band.

Figure 7 shows an example of CDAD for the valence band of graphene-intercalated indenene, i.e., a monolayer of In atoms sandwiched between a SiC(0001) substrate and a monolayer of graphene on SiC. Indenene has been demonstrated to be a two-dimensional quantum spin Hall insulator [25], with the graphene capping acting as efficient protection against environmental influences [26]. CDAD provides information on the orbital angular momentum and hence the Berry curvature in the band structure of this topological insulator [27]. Figures 7a,b display the momentum patterns measured with right and left circularly polarized light, respectively, at a photon energy of 230 eV.

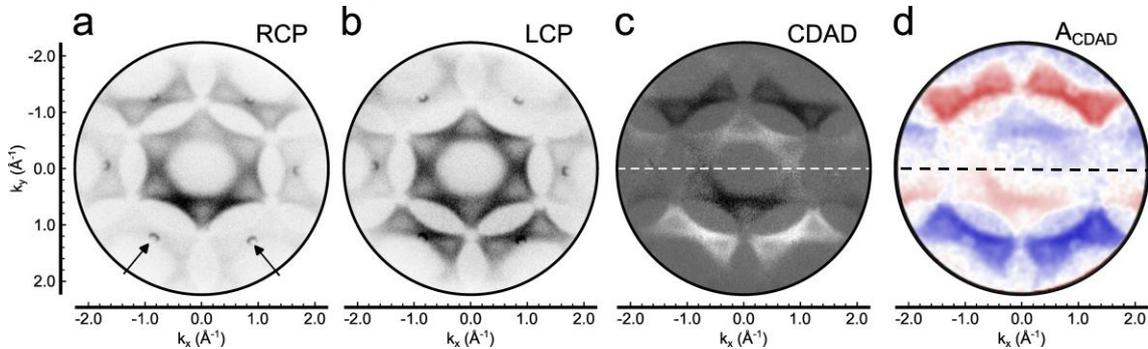

**Fig. 7.** CDAD for the valence band of indenene intercalated into the graphene / SiC interface. (a,b) Constant energy maps measured at a photon energy of 230 eV with circular right and left polarized light, respectively. (c) Resulting CDAD in greyscale and (d) CDAD asymmetry colorized showing textbook-like dichroism with the incidence direction of the synchrotron radiation as mirror axis (horizontal).

The graphene K points are clearly visible (arrows in (a)) and overlayed with the In 5*p* bands. A pronounced dichroism contrast from top to bottom is visible, which reverses with switching the photon helicity (cf. (a) and (b)). The circular dichroism is quantified in (c) as the difference signal and in (d) as asymmetry, displayed in blue and red for positive and negative asymmetry. The CDAD asymmetry reaches almost 100% and exhibits a textbook-like antisymmetry with respect to the horizontal mirror plane running through the centre (dashed lines). Note that the use of soft X-ray photoemission with its enhanced probing depth has allowed us to probe the indenene band structure through its graphene capping, i.e., in a device-like geometry. This demonstrates the high potential of this momentum microscope for *in-operando* spectroscopy of functional nanostructures.

As electronic systems prototypical for 3D behavior, we have chosen Au and Cu crystals with the (111) surface orientation. The procedure for the determination of the Fermi surface was explained in Sec. 3.2. Here we show the rich CDAD structure of the *sp* valence band patterns, measured with circularly polarized soft X-rays at I09. A selection of $k_x$-$k_y$ sections at the Fermi energy recorded at different photon energies is displayed in Fig. 8. The first and second rows show the intensity patterns for Au and Cu (sum $I_{RCP}$ + $I_{LCP}$) the third and fourth rows the corresponding circular dichroism textures, respectively.

We adopt the terminology from dHvA measurements, as introduced in the pioneering work of Shoenberg [19,20]. Figure 8p shows half of the Fermi surface, starting from the 'belly' down to the 'neck' toward the next BZ, with planes (i-iv) indicating the positions of the $k_x$-$k_y$ sections. The cut (i) at the belly is shown in Fig. 6a. In the first column of Fig. 8 we show plane (ii), where the central Fermi surface touches the necks to the adjacent BZs. The intensity patterns (a) for Au and (f) for Cu show how the central sixfold-warped hexagon touches the hexagons of three next BZs that are 120° apart. Due to the larger lattice constant of Au, its BZ is smaller. Hence, for Au (a) the momentum microscope captures a larger fraction of the neighboring BZs than for Cu (f). The second, third and fourth columns show three sections near cut (iii), where the central Fermi surface contracts. In Figs. 8b,g we see threefold-warped patterns in the centre, which contract further in (c,h) and (d,i). Finally, in the fifth column we probe the neck position (iv). It shows a clear ring-shaped Fermi surface cut in the center for Cu (j), whereas for Au the center appears as bright point (e). In sections (e,j) the six outer hexagons of the adjacent BZs touch each other.

The CDAD patterns exhibit a rich structure, resulting from interference effects. In sections (k) and (o) the dichroism shows a simpler plus/minus feature for the central Fermi surface. However, the fine structure of CDAD along the prominent narrow band features changes with $k_z$ position. This effect is strongest for the central band feature in the contraction region, Figs. 8 l,m,n,q,r,s. In particular (l) shows a complicated pattern, which is also reflected by the intensity pattern (b) that reveals more than one band. This is likely due to different final state bands leading to the interference effect. We note that a detailed discussion of CDAD of the Au(111) Shockley surface state is given in Ref. [28].

Remarkably, the 'diffuse' intensity background outside of the band features shows a strong dichroism with a pronounced texture. This effect is much stronger for Au (third row), presumably due to the higher scattering cross section of the heavy Au atoms in comparison with Cu (bottom row). Photoemission calculations will shed light into this phenomenon.

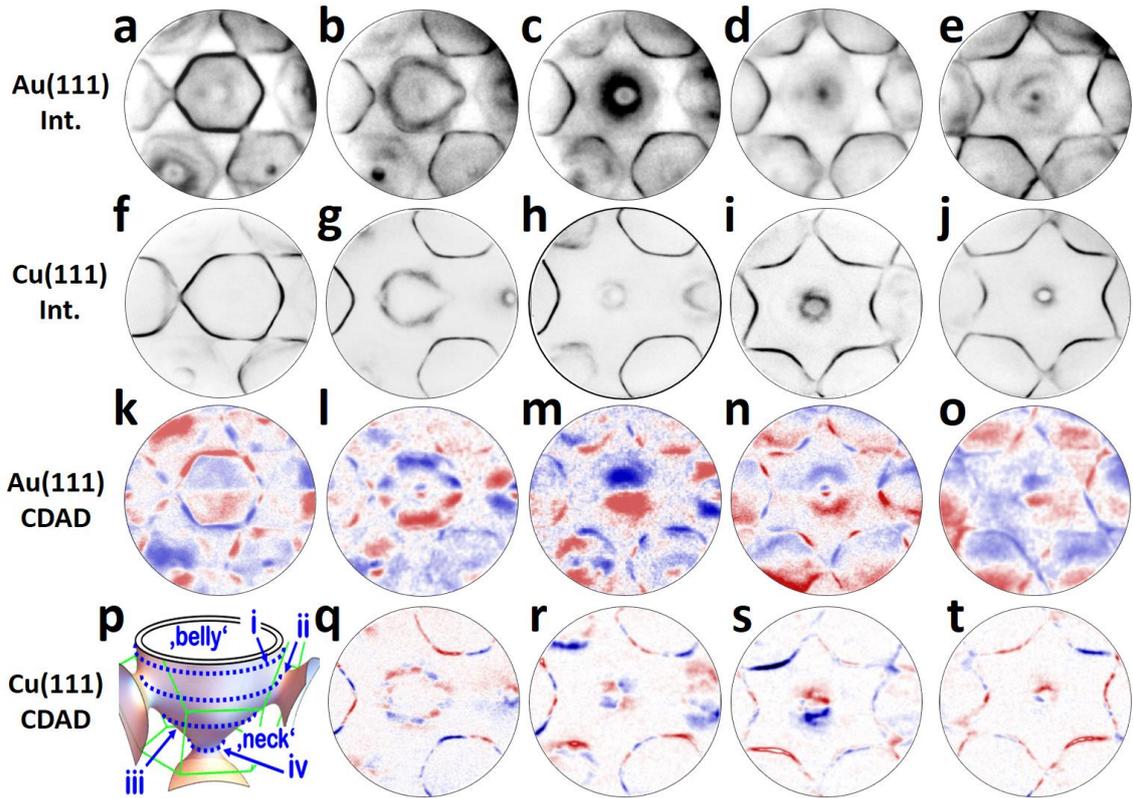

**Fig. 8.** $k_x$-$k_y$ intensity patterns for Au(111) (a-e) and Cu(111) (f-j) at the Fermi energy for different photon energies between 520 and 720 eV, corresponding to different values of $k_z$. (k-o) and (q-t) show the corresponding circular dichroism (CDAD) textures. (p) View of the lower half of the theoretical Fermi surface with indicated planes (i-iv) of the $k_x$-$k_y$ sections (see text). [(p) after Ref. [22]].

### 3.4 Full-field photoelectron diffraction and circular dichroism texture

Parallel imaging of large *k*-fields-of-view enables a new, highly efficient approach to photoelectron diffraction (PED; in the X-ray range also termed XPD), which requires high angular resolution (our instrument reaches 0.03°) and high energy resolution (usually given by the photon bandwidth) and can thus fully exploit the capabilities of the momentum microscope in an ideal way. Full-field diffractograms reveal complex patterns of intersecting Kikuchi bands [29]. In our instrument, PED analysis can be combined with valence-band mapping under identical conditions just by tuning from the core-level signals to the valence band. Combining PED and valence-band mapping in a single experiment gives access to the interplay of geometric and electronic structures, e.g., when crossing a structural phase transition. Imaging PED is particularly attractive because of its unprecedented acquisition speed. For intense core levels, such diffractograms can be watched in real time with a frame rate of 1 Hz.

Alongside with the possibility to provide a detailed structural analysis, PED patterns also have practical aspects: They provide an alternative metric for *k*-space and show distinct directions in real space, e.g., the azimuthal orientation and precise position of the sample surface (important for samples with different domains). This enables rapid alignment of sample and microscope settings. Examples of diffractograms for the Ge 3*d* and 3*p* core levels are shown in Fig. 9. The diffraction kinematics is determined by the final state of the photoelectrons *inside* the material $E_{final}$ = hν - $E_B$ + $V_0$, where $E_B$ is the binding energy of the core level and $V_0$ the inner potential.

The conventional framework to analyze photoelectron diffraction patterns are real-space polar coordinates [30], whereas in our instrument we capture PED patterns on a momentum scale. At hard X-ray energies the diffractograms are dominated by Kikuchi bands [31], which contain both, *k*-space and real-space information. The widths of Kikuchi bands correspond to multiples of reciprocal lattice vectors; hence they provide a metric for *k*-space coordinates [32]. On the other hand, the center lines and crossing points of bands correspond to projected lattice planes and atom rows along high-symmetry directions of the crystal in real space. The filigree fine structure of diffractograms in the hard X-ray range originates from the short photoelectron wavelength combined with the large IMFP [18]. For photoelectrons with $E_{final}$ of several keV, $10^5$ to $10^6$ atoms contribute to the diffraction process. This is the classical *Kikuchi regime* as perfectly understood in electron microscopy [33]. In this range we gain true bulk information, because the surface contribution is negligible. Many-beam dynamical Kikuchi-band theory using the Bloch-wave approach [34] yields excellent agreement with experiments (for an overview, see [35]).

Photoelectron energies in the 100 eV range correspond to the minimum of the IMFP curve. Here, PED obtains almost pure surface information with the bulk structure being essentially invisible. In this range, PED has been used to analyze surface relaxation and reconstruction and adsorbate structures, in particular adsorption sites (see, e.g., review articles [36,37]). Due to the short IMFP the nearest-neighbor atoms contribute dominantly to the scattering processes. This regime is commonly termed the *holographic regime* because it is possible to retrieve the real-space atom configuration from the diffraction pattern [38]. Matsushita et al. [39] introduced the concept of 'quasi Kikuchi bands', which is based on coherent scattering at a single atom row or atom plane.

Full-field imaging PED at low energies in the 100 eV range has up to now only been performed using the HEXTOF momentum microscope [40] at the free-electron laser FLASH. A related type of structural information has been obtained with the same instrument by imaging the *k*-pattern of photoelectrons from molecular orbitals, termed orbital tomography [41].

At energies between 100 eV and 1 keV, the so-called immersion ratio $\sqrt{[E_{kin}/(E_{kin}+eU_{extr})]}$ of the extractor lens varies from 0.1 to 0.3. The voltage $U_{extr}$ between sample and extractor accelerates the electrons to high energies, typically 10 keV. As a consequence, the recorded range of polar angles shrinks from 30° at 106 eV to 11° at 1036 eV. The dependence of parallel momentum on polar angle θ and initial kinetic energy $E_{kin}$ is given by $k_\| = 0.512 \sin\theta \sqrt{E_{kin}}$. Since the approximation $E_{kin} \ll eU_{extr}$ is no longer valid, the lateral magnification of the objective lens varies slightly and the *k*-field-of-view increases, here from 5.0 Å$^{-1}$ (Fig. 9a) to ~6 Å$^{-1}$ (Figs. 9e,g). The properties of cathode lenses for different kinetic energies were studied in detail by Bauer [42] and Tromp et al. [43]. In the future, a modified front lens will give access to even larger *k*-fields-of-view [44]; here we do not go into further details.

The PED data shown in Fig. 9 cover the energy range from close to the minimum of the IMFP curve into the regime of bulk sensitivity. Between $E_{final}$ = 106 and 1036 eV (Fig. 9a-e) the IMFP for Ge increases from 0.5 nm to 2.2 nm [18]. Nevertheless, all patterns are rich in details even at the lowest energies. Figs. 9f,h show calculations using the Bloch-wave approach [34]. Comparison of (f) with the experimental patterns (e) reveals good agreement both in the appearance of the Kikuchi band marked by dashed lines and the fine structure in the center region. The simulation in (h) shows a larger $k$-field-of-view, illustrating crossing Kikuchi bands. The dashed circle indicates the diameter of the measured patterns.

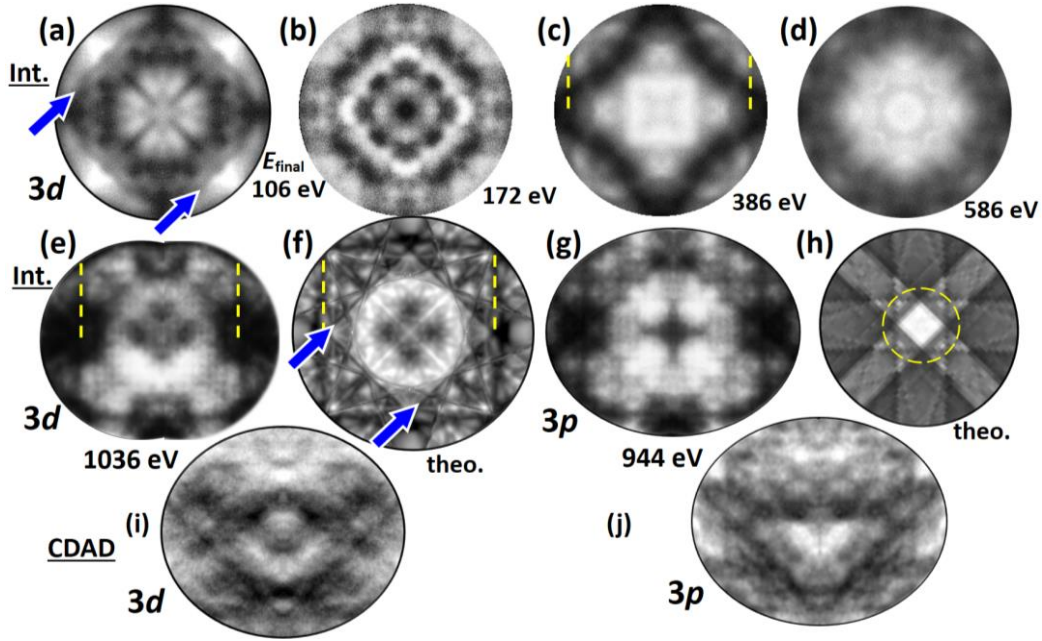

**Fig. 9.** Photoelectron diffraction and circular dichroism in Ge 3$d$ and 3$p$ core-level photoemission from a Ge(001) surface. (a-d) $k_x$-$k_y$ momentum images for Ge 3$d$ at different final-state energies as stated in the panels. (e) Same for $E_{final}$ = 1036 eV with corresponding calculated diffractogram (f). The arrows in (f) mark the same diamond feature as visible in (a). (g) Same for the Ge 3$p$ core level at $E_{final}$ = 944 eV. (h) Calculated pattern with a larger diameter. Bottom row: Circular dichroism signal for Ge 3$d$ (i) and 3$p$ (j).

The strongly structured PED patterns for kinetic energies as low as 106 and 172 eV (a,b) are surprising. More systematic calculations will elucidate the reason for the almost 'bulk-like' appearance of the patterns in the first row. We see clear borders of the central Kikuchi band, dashed lines in (c), and the diamond shaped feature marked by arrows in (a,f). Due to the immersion ratio effect, pattern (a) is expanded in comparison with (f). The fine structure of the central region exhibits a strong energy dependence in the sequence (a-d).

At h$\nu$ = 1050 eV the Ge 3$d$ and 3$p$ patterns show a significant circular dichroism in the angular distribution CDAD, Figs. 9i,j. The origin of this effect is discussed in Ref. [45]. The CDAD requires a non-coplanar geometry [46] of the helicity vector of the incoming photon beam, the photoelectron momentum and the electronic quantization axis, here the surface normal. As a consequence of this symmetry property, the CDAD vanishes in the horizontal mirror plane and the regions above and below the mirror plane show an anti-symmetric CDAD pattern. The dichroism texture differs strongly from the corresponding intensity patterns, Figs. 9e,g. CDAD is a matrix element effect and originates from the local excitation at the emitter atom. However, the Kikuchi diffraction process strongly modifies the CDAD, leading to complex asymmetry patterns.

## 4. Summary and outlook

We have presented a novel photoelectron momentum microscope installed at the soft X-ray branch of beamline I09 of the Diamond Light Source. The key element of the momentum microscope is a large single hemispherical spectrometer with a ToF analyzer behind the exit slit. The beamline covers a photon energy range from 110 eV to 2 keV, allowing both, surface and bulk sensitive ARPES measurements. Circular polarization is available for hν > 210 eV, for measurements of the circular dichroism in valence band mapping (CD-ARPES). First experiments show energy and momentum resolutions of 10.2 meV and 0.010 Å$^{-1}$ at a pass energy of 10 eV and sample temperature of 28 K. At lower pass energies, a resolution of 4.2 meV was achieved in a laboratory experiment prior to installation at the beamline. The large angular filling of the entrance lens and the hemisphere allows $k$-fields-of-view > 6 Å$^{-1}$, which give access to part of the first repeated Brillouin zone (BZ) for many materials. A focused and monochromatized He lamp is used for off-line measurements at hν = 21.2 eV (He I) and 40.8 eV (He II). The momentum microscope is integrated into a versatile sample handling and preparation system, including equipment for surface preparation and analytics (sputtering, annealing, LEED).

The ToF analyzer behind the exit slit of the hemisphere ('ToF booster') is triggered by the bunch marker of the storage ring. DLS provides a photon pulse train with a period of 2 ns. Due to its time resolution of ~200 ps, the ToF analysis allows the decomposition of the energy band transmitted by the hemispherical analyzer into several (ideally 10) time slices. Due to limitations at the edges of the time gap, only about 5 time slices are actually available for analysis. In addition to this factor of 5 due to the 'multiplexing', another gain factor results from the fact that the hemispherical analyzer can be operated with about 10 times higher pass energy than in the operating mode without ToF analysis. The higher pass energy corresponds to higher analyzer transmission, with the gain factor depending on the size of the photon footprint relative to the accepted area on the sample surface. We estimate a total gain factor of 15 when the ToF booster is turned on.

In addition to parallel imaging ARPES and CD-ARPES, the entrance lens can be tuned for energy filtered real space imaging (X-PEEM). First experiments with lithographic test samples showed a lateral resolution of about 300 nm. Simulations [41] suggest resolutions in the <100 nm range, provided that the high photon flux density in the analyzed region is sufficiently high. The large $k$-field-of-view is favorable for recording photoelectron diffraction (PED) patterns and their circular dichroism texture (CD-PED).

Examples are given for 2D band mapping of bilayer graphene and the circular dichroism texture of the two-dimensional quantum spin Hall insulator indenene. The three-dimensional band structure and CDAD texture of the valence bands of gold and copper were measured in the photon energy range between 520 and 720 eV. Surprisingly, the results for these two prototypical bulk metals with $sp$ valence bands revealed some unexpected details that need to be further elucidated by ongoing experiments in addition to theoretical calculations. As an example of full-field photoelectron diffraction we showed a series of PED patterns for the Ge 3$p$ and 3$d$ core levels in the range of final state energies between 106 eV and 1 keV. Despite their small inelastic mean free path at low kinetic energies the PED patterns show a lot of fine structure that varies strongly with kinetic energy. Calculations of full-field imaging photoelectron diffractograms at low kinetic energies in the 100 eV range are highly desirable.

With its unique capability for massive parallel *k*-space mapping at competitively high energy resolution our hybrid momentum microscope represents the next generation of photoemission instrumentation. Its use with soft X-ray excitation and the concomittant higher photoelectron mean free path allows for high-precision 3D bulk band mapping, but also provides access to buried interfaces and nanostructures and facilitates in-operando spectroscopy of real devices. Furthermore, in combination with linear and circular dichroism, momentum microscopy has the potential to boost photoemission from a mere spectroscopy of energy eigenvalues to the tomography of the associated wave functions, i.e., providing full information on the complex Bloch eigenfunctions (and all related quantum properties, such as Berry curvature or quantum geometric tensor). Finally, our momentum microscope is ideally suited to be combined with 2D spin filtering to realize a complete quantum measurement of the band structure. This is planned for future work.

## Acknowledgments

We thank the Diamond Light Source for providing beamtime. Sincere thanks go to Dave McCue for excellent system design and valuable technical support. We thank Aimo Winkelmann (U Krakóv) for support with the PED simulations, Surface Concept GmbH (Mainz) for support with the delayline detector, Phil King (U St Andrews) for the bilayer graphene, Jonas Erhardt and Cedric Schmitt (both at U Würzburg) for the graphene-intercalated indenene samples, and Simon Moser (U Würzburg) for helpful discussions on the circular dichroism. This work was funded by Deutsche Forschungsgemeinschaft DFG through projects CL 124/16-1, SCHO 341/16-1 and the Würzburg-Dresden Cluster of Excellence on Complexity and Topology in Quantum Matter 'ct.qmat' (EXC2147, project ID 390858490). Further support was also provided by Diamond.